\documentclass[aps,prb,twocolumn,superscriptaddress,showpacs,floatfix]{revtex4}

\usepackage{graphicx}
\graphicspath{{figs/}}
\begin{document}
\title{First principle study of the thermal conductance in graphene nanoribbon with vacancy and substitutional silicon defect}
\author{Jin-Wu~Jiang}
    \altaffiliation{Electronic address: phyjj@nus.edu.sg}
    \affiliation{Department of Physics and Centre for Computational Science and Engineering,
             National University of Singapore, Singapore 117542, Republic of Singapore }
\author{Bing-Shen~Wang}
    \affiliation{State Key Laboratory of Semiconductor Superlattice and Microstructure and Institute of Semiconductor, Chinese Academy of Sciences, Beijing 100083, China}
\author{Jian-Sheng~Wang}
    \affiliation{Department of Physics and Centre for Computational Science and Engineering,
                 National University of Singapore, Singapore 117542, Republic of Singapore }

\date{\today}
\begin{abstract}
The thermal conductance in graphene nanoribbon with a vacancy or silicon point defect (substitution of C by Si atom) is investigated by non-equilibrium Green's function (NEGF) formalism combined with first-principle calculations density-functional theory with local density approximation. An efficient correction to the force constant matrix is presented to solve the conflict between the long-range character of the {\it ab initio} approach and the first-nearest-neighboring character of the NEGF scheme. In nanoribbon with a vacancy defect, the thermal conductance is very sensitive to the position of the vacancy defect. A vacancy defect situated at the center of the nanoribbon generates a saddle-like surface, which greatly reduces the thermal conductance by strong scattering to all phonon modes; while an edge vacancy defect only results in a further reconstruction of the edge and slightly reduces the thermal conductance. For the Si defect, the position of the defect plays no role for the value of the thermal conductance, since the defective region is limited within a narrow area around the defect center.

\end{abstract}

\maketitle

\pagebreak

\section{introduction}
Defect engineering is a powerful approach to generate nano-materials with plenty of novel properties\cite{TerronesM}. For instance, the defect induced magnetism in graphene was predicted\cite{YazyevOV,PalaciosJJ}, and a patterned defect can be used to open up a band gap in graphene by breaking the sublattice symmetry of the honeycomb structure\cite{AppelhansDJ}. The defects formation in graphene can be observed directly by high-resolution transmission electron microscopy (TEM)\cite{HashimotoA}. The formation and annealing of Stone-Wales and other defects were observed by transmission electron aberration-corrected microscope monochromated (TEAM 0.5) within 1~{\AA} resolution at an acceleration voltage of only 80 kV that is low enough to protect graphene samples from being destroyed\cite{MeyerJC}. These techniques for observation of defects in atomic-scale serve as a fundamental support for the nanoengineering of defects in graphene. The engineering of defects on the two-dimensional graphene sheet has been proposed theoretically in 2008 by Lusk {\it et~al.} using first-principle calculations\cite{Lusk2008}. New carbon allotropes can be fabricated by periodically arrangement of basic defect building blocks such as Stone-Wales defects, inverse Stone-Wales defects, vacancy defects, and etc\cite{Lusk2008}. The inverse Stone-Thrower-Wales defect was also suggested as a building block for monolithic nanoengineering on graphene\cite{Lusk2010}. Quite recently, a one-dimensional topological defect in graphene is realized in the experiment\cite{LahiriJ}. This extended defect exhibits similar property as a metallic wire and may form building blocks for various one-dimensional nano devices. In present paper, we will focus on the vacancy defect and substitutional Si defect in graphene nanoribbon.

In the quasi-one-dimensional nanoribbon structure, the two edges undergo a reconstruction and play an important role in determining many properties\cite{KoskinenP2008,KoskinenP2009}. The TEAM 0.5 design was implemented to observe the edge reconstruction and capture the complex behavior of a atom at the edge of the graphene nanoribbon\cite{GiritGO}. As the edge region of a nanoribbon has very different properties from the inner region, an edge defect should behavior quite differently from a defect inner the nanoribbon. Particularly, we are interested in how the edge and inner point defects behavior distinctly on the phonon thermal conductance in the graphene nanoribbon. The thermal transport capability is important in many functional nano devices. High thermal conductivity is a key property for high performance electronic devices, while extremely low thermal conductivity is crucial for thermoelectric materials. For graphene sheet, a high thermal conductivity was discovered experimentally\cite{Balandin}. Various theoretical approaches have been applied to study the thermal transport in graphene system. The direct non-equilibrium molecular dynamics simulation can be used to mimic the experimental conditions to study the thermal transport in graphene\cite{YangN2008,HuJN,YangN2009,JiangJW2010}. The Bolzmann equation method is suitable for the study of the diffusive thermal transport. The non-equilibrium Green's function (NEGF) approach\cite{YamamotoT,MingoN,WangJS2008} and a molecular dynamics simulation with quantum heat baths\cite{WangJS2009} are both proper methods for the study of quantum thermal conductance. In our calculation, we apply the NEGF approach.

In this paper, we study the phonon thermal conductance in graphene nanoribbon by using the NEGF approach combined with the first-principle calculation. The {\it ab initio} calculation gives a force constant matrix with long-range interaction while the NEGF approach requires first-nearest-neighboring (FNN) interaction in the calculation of the surface Green's function. This conflict may be improved by using an enlarged unit cell. However, the enlargement of unit cell distinctly increases the simulation cost, which is even worse in {\it ab initio} calculation. We thus propose an efficient correction to the force constant matrix to effectively renormalize the long-range interaction components into the FNN part. We then investigate vacancy and Si point defects in the nanoribbon. The position of the vacancy defect plays an important role. An edge vacancy defect results in a further reconstruction of the edge of the nanoribbon thus only slightly affects the thermal conductance. The inner vacancy defect causes the defective area to become saddle-shaped, which considerably reduces the thermal conductance. All phonon modes are strongly scattered by this saddle-like surface in the nanoribbon. In the case of Si defect, the thermal conductance shows no difference for edge or inner point defects. We also study phonon modes in the nanoribbon and find that there are some phonon modes localized at the edge of the nanoribbon. The number of edge modes is determined by the number of C-C bonds at the edge of the nanoribbon. These edge modes are of much higher frequency, since the edge C-C bonds are shorter and stronger after reconstruction during structure relaxation. The edge vacancy or Si defects cause a significant downshift of a particular edge mode localized around this defective area.

\section{combination of Siesta and NEGF}
In this section, we concentrate on the combination of the NEGF approach and the first-principle calculation. This is nontrivial, because the first-principle calculation gives a force constant matrix with long-range interaction, while what the NEGF requires is a force constant matrix of FNN interaction. We have to take care of the conflict between the long-range and FNN characters.

In the {\it ab initio} calculation, we use the SIESTA package\cite{SolerJM} to optimize the structure of the graphene nanoribbon and extract the force constant matrix. The local density approximation is applied to account for the exchange-correlation function with Ceperley-Alder parametrization\cite{CeperleyDM} and the double-$\zeta$ basis set orbital (DZ) is adopted. During the conjugate-gradients optimization of the nanoribbon, the maximum force on each atom is smaller than 0.01 eV/\AA. A mesh cut off of 320 Ry is used. Periodic boundary condition is applied in the heat transport direction, while the nanoribbon is free in the two vertical directions by introducing large vacancy space. We employ Gamma point $k$ sampling in the periodic direction.

As discussed in previous works\cite{MingoN2008}, the original force constant matrix $K_{0}$ from SIESTA violates the rigid translational symmetry (RTS) and rigid rotational symmetry for the lattice structure. RTS is the origin of the zero frequency for three acoustic phonon modes at $\Gamma$ point, while the rigid rotational symmetry is the root for the zero frequency rotational phonon modes in cylinder-like structures such as carbon nanotubes. In the planar graphene sheet, it is important to ensure that the force constant matrix satisfies the RTS, so that the three acoustic phonon modes have zero frequency. The RTS can be enforced by a renormalization of $K_{0}$\cite{MingoN2008}. We refer to this new force constant matrix as $K$. The $K$ matrix can be written into following explicit form:
\begin{eqnarray}
K & = & \left(\begin{array}{ccccc}
K_{00} & K_{01} & K_{s} & 0 & K_{10}\\
K_{10} & K_{00} & V^{LC} & 0 & K_{s}^{\top}\\
K_{s}^{\top} & V^{CL} & K_{C} & V^{CR} & 0\\
0 & 0 & V^{RC} & K_{00} & K_{01}\\
K_{01} & K_{s} & 0 & K_{10} & K_{00}\end{array}\right),
\label{eq_K}
\end{eqnarray}
where we have used labels L/C/R to denote three different regions: left lead, center, and right lead. The division of a whole system into three different regions is conventional in NEGF approach\cite{YamamotoT,MingoN,WangJS2008}. The two matrices at the right top and left bottom corners are resulting from the periodic boundary condition in the heat transport direction. $V^{LC}$ and $V^{RC}$ are coupling between leads and center. The force constant matrix for left/right leads have been written in a periodic form, since the leads are the repetetion of the smallest unit cell. The dimension of $K_{00}$, $K_{01}$ and $K_{10}$ is the total degree of freedom in the unit cell. $K_{00}$ is the force constant matrix for each unit cell and $K_{01}$ is the coupling between two neighboring unit cells. Essentially, they are used to reconstruct the force constant matrix of the leads in the NEGF scheme:
\begin{eqnarray}
K_{L} & = & \left(\begin{array}{ccccc}
\ddots & \ddots & \ddots & 0 & 0\\
0 & K_{10} & K_{00} & K_{01} & 0\\
0 & 0 & K_{10} & K_{00} & K_{01}\\
0 & 0 & 0 & K_{10} & K_{00}\end{array}\right).
\label{eq_KL}
\end{eqnarray}
We show the result of left lead and the right lead is analogous. The most important physical quantity in the NEGF approach is the surface Green's function of the lead, which is determined by the $K_{L}$. The $K_{L}$ is a semi-infinite matrix as the lead is semi-infinite. As a result, it is impossible to calculate the surface Green's function from an inverse of the force constant matirx by the expression $g_{L/R}=\left[\left(\omega+i\eta\right)^{2}-K_{L/R}\right]^{-1}$. An efficient iterative method has been developed by Sancho et al to solve the surface Green's function numerically\cite{SanchoMPL}. This approach takes the advantage of the periodicity of the lead. It also requires that each unit cell only interacts with its two FNN unit cells. We refer to this requirement as the FNN property of the NEGF approach. After obtaining the $g_{L/R}$, the retarded Green's function for the center region can be calculated by:
\begin{figure}[htpb]
  \begin{center}
    \scalebox{1.0}[1.0]{\includegraphics[width=8cm]{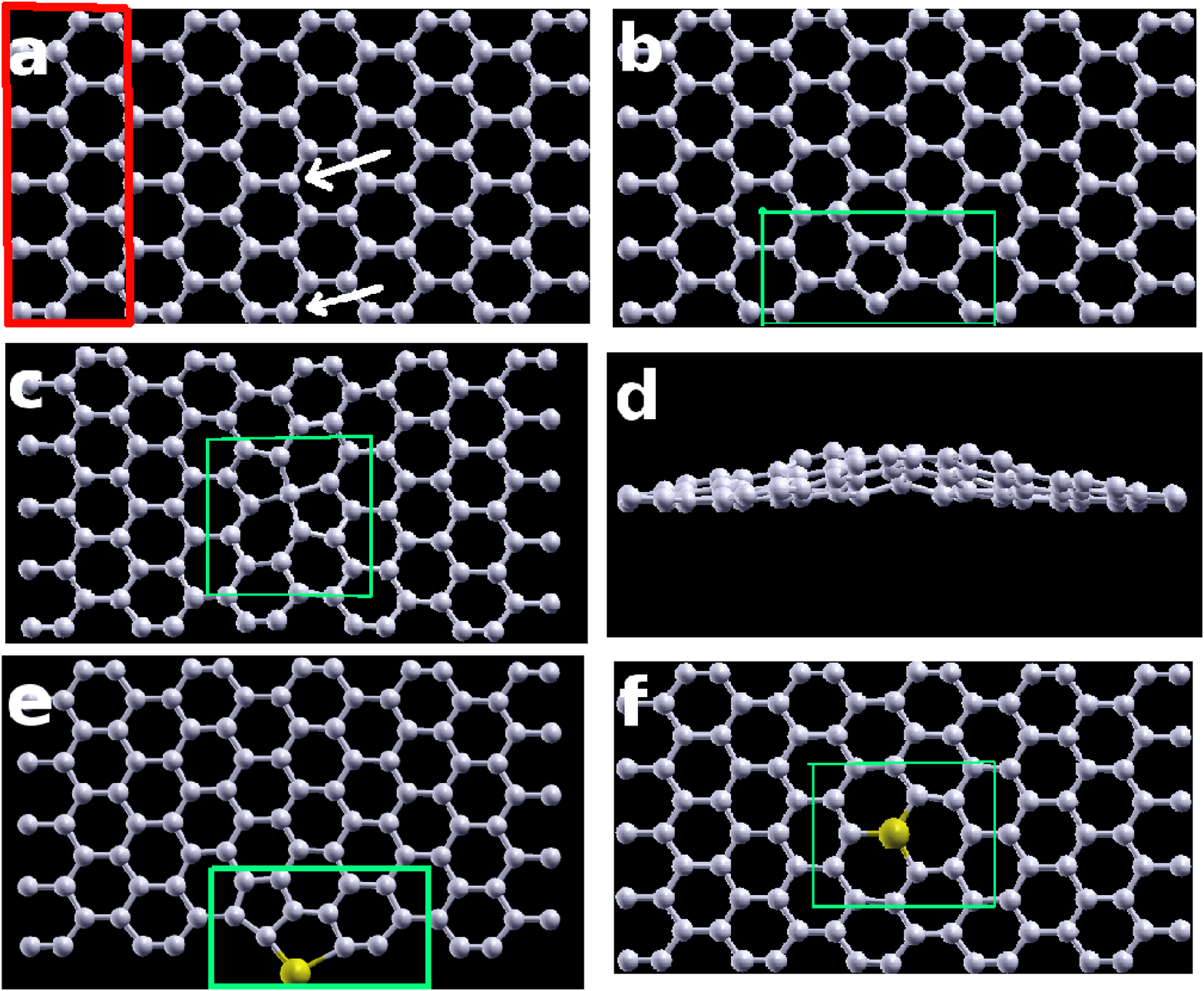}}
  \end{center}
  \caption{(Color online) Optimized configurations for graphene nanoribbon. (a). Pristine nanoribbon. The unit cell is highlighted by a thick box (red online). (b). Vacancy defect at edge. (c), (d) Vacancy defects inner nanoribbon. (e). Si defect at the edge. (f). Si defect inner nanoribbon. Two arrows in (a) denotes the position for vacancy or Si defects. The defective regions are highlighted by a thin box (green online).}
  \label{fig_cfg}
\end{figure}
\begin{eqnarray}
G^{r} & = & \left[\left(\omega+i\eta\right)^{2}-K_{C}-\Sigma_{L}-\Sigma_{R}\right]^{-1},
\end{eqnarray}
where the self-energy for the lead is
\begin{eqnarray}
\Sigma_{L} & = & V^{CL}g_{L}V^{LC}.
\end{eqnarray}
The transmission for phonon is given by the Caroli formula:
\begin{eqnarray}
T[\omega] & = & Tr\left(G^{r}\Gamma_{L}G^{a}\Gamma_{R}\right),
\end{eqnarray}
where $G^{a}=\left(G^{r}\right)^{\dagger}$ is the advanced Green's function and $\Gamma_{\alpha}=-2Im\Sigma_{\alpha}^{r}$. Finally, the phonon thermal conductance can be obtained by Landau formula: 
\begin{eqnarray}
\sigma_{ph} & = & \frac{1}{2\pi}\int d\omega\hbar\omega T_{ph}[\omega]\left[\frac{\partial n(\omega,T)}{\partial T}\right].
\end{eqnarray}

Now we keep track of the RTS in the whole NEGF scheme as shown above. Overall, there are two direct consequences from the RTS. Firstly, the diagonalization of $K$ should give three zero-frequency acoustic phonon modes. Secondly, corresponding to these three modes, the transmission function from NEGF should be 3 at $\omega=0$, i.e $T[\omega=0]=3$. In the above, as the $K$ matrix in Eq.~(\ref{eq_K}) satisfies the RTS, we can correctly achieve zero frequency for the three acoustic phonon modes. So the first consequence of the RTS is obtained.
\begin{figure}[htpb]
  \begin{center}
    \scalebox{1.0}[1.0]{\includegraphics[width=8cm]{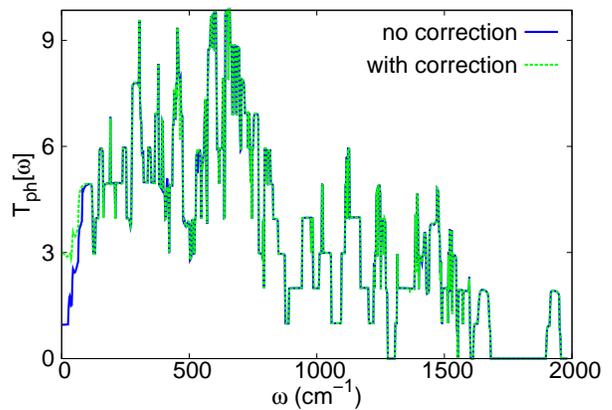}}
  \end{center}
  \caption{(Color online) The transmission function with and without correction to the force constant matrix for pure graphene nanoribbon.}
  \label{fig_trans_correction}
\end{figure}
 We now check the second consequence. As can be seen from the first line of $K$, the RTS results in following relation
\begin{eqnarray}
\left(K_{00}+K_{01}+K_{s}+K_{10}\right)I_{dof} & = & 0,
\end{eqnarray}
where $I_{dof}$ is a unit row vector with dimension as the total degree of freedom in unit cell. $K_{s}$ corresponds to the long-range interaction beyond the FNN. The nonzero $K_{s}$ is important to keep the whole $K$ matrix satisfying RTS. Because of $K_{s}\not=0$, it can be seen from the first line of $K_{L}$ in Eq.~(\ref{eq_KL}) that RTS has been broken in the lead, as $
\left(K_{00}+K_{01}+K_{10}\right)I_{dof} \not= 0$. It leads to $T[\omega=0]\not=3$ as shown in Fig.~\ref{fig_trans_correction} by the solid line (blue online). This is the conflict between the long-range character of the first-principle calculation and the FNN property of the NEGF.

A straightforward technique to improve this conflict is to enlarge the unit cell of the leads to obtain smaller value of $K_{s}$\cite{SanchoMPL}. One can get $K_{s}=0$ for large enough unit cell. However, larger unit cell leads to the increase of the size of whole system, thus the calculation cost is increased and even more difficult in the first-principle calculation\cite{TanZW}. Another possible way to solve the conflict is doing correction to $K$ matrix, while the unit cell kept unchanged. There is no increase of calculation cost in this method since the size of the unit cell is the same. We can use the smallest unit cell in the nanoribbon as shown in Fig.~\ref{fig_cfg}~(a). This correction is realized in following two steps.

(a). As discussed above, we enforce the RTS to the force constant matrix. This will give us the $K$ matrix with translational symmetry fulfilled. We stress that $K$ has a nonzero matrix $K_{s}$.

(b). We set $K_{s}=0$ in $K$ matrix, resulting in a new force constant matrix $K'$.

\begin{figure}[htpb]
  \begin{center}
    \scalebox{1.0}[1.0]{\includegraphics[width=8cm]{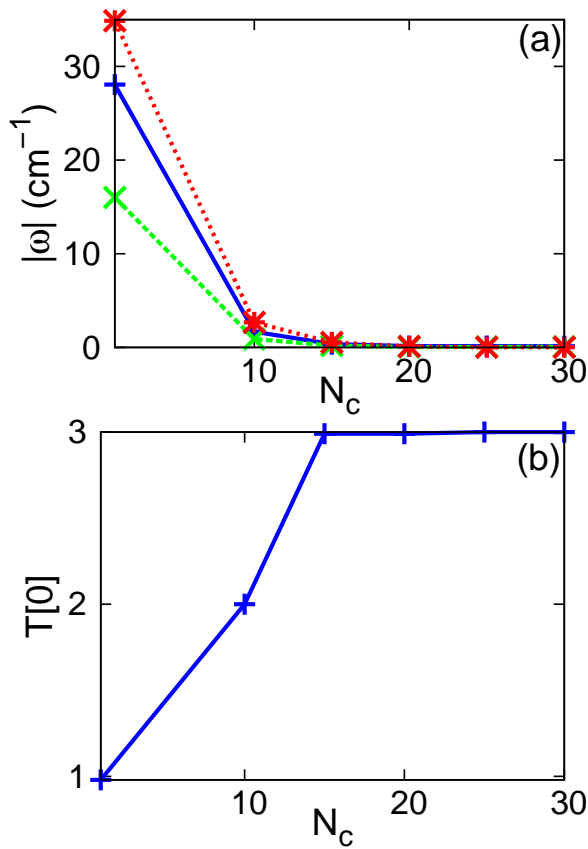}}
  \end{center}
  \caption{(Color online) The convergence of the correction to force constant matrix. (a). Frequency of three acoustic modes v.s correction steps $N_{c}$. (b). The transmisstion function at $\omega=0$ from NEGF approach v.s correction steps.}
  \label{fig_correction_step}
\end{figure}
Obviously, $K'$ from step (b) violates the RTS. We need to go back to step (a) to enforce RTS to $K'$ matrix and will obtain a new force constant matrix $K$, which satisfies the RTS and has a new nonzero $K_{s}$. This nonzero matrix $K_{s}$ is smaller than the original one. After many cycles of these two steps, we obtain a force constant matrix $K$ with RTS fulfilled and $K_{s}\approx 0$. So the FNN property of the NEGF can also be satisfied. Essentially, the long-range component, $K_{s}$, is renormalized into the FNN part by the correction. As shown in Fig.~\ref{fig_trans_correction} by dot line (green online), after 20 cycles of steps (a) and (b), the transmission function from NEGF gives a correct result, i.e $T[\omega=0]\approx 3$. The correction has a nice property that it only affects the low-frequency region of the transmission function, indicating that the RTS is enforced without affecting higher frequency region. The convergence of this correction is demonstrated in Fig.~\ref{fig_correction_step}. It shows that after 10 cycles, zero frequency is achieved for the three acoustic phonon modes from $K$ and the $T[\omega=0]\approx 3$ from NEGF approach. The correction to the $K$ matrix can successfully combine the long-range SIESTA with the FNN NEGF sheme.

Before continuation of further numerical calculation, we give a brief summation for this part. (1). We are discussing the conflict between the long-range character of SIESTA and the FNN property of NEGF. The conflict is explicitly manifested by the RTS in the graphene nanoribbon. (2). We solve this conflict by effectively renormalizing the long-range component, $K_{s}$, into other FNN components in the $K$ matrix.

\section{calculation results and discussion}
\begin{figure}[htpb]
  \begin{center}
    \scalebox{1.0}[1.0]{\includegraphics[width=8cm]{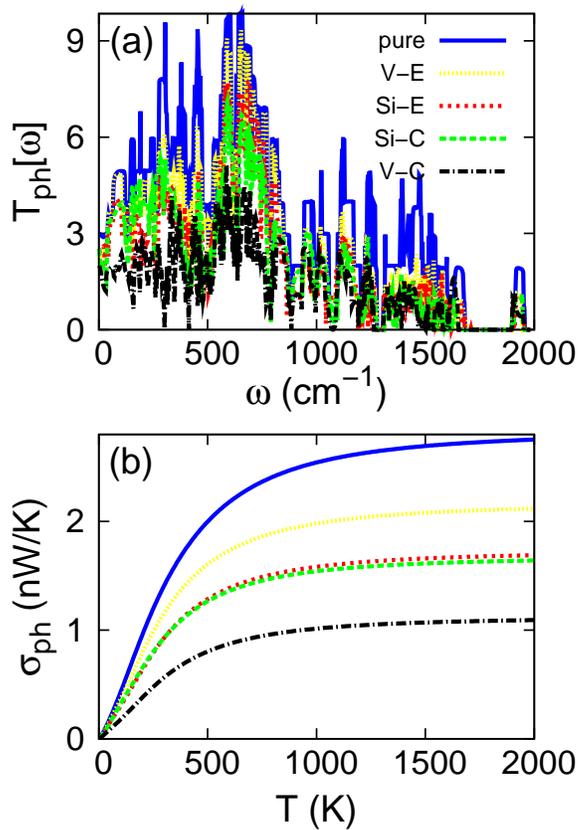}}
  \end{center}
  \caption{(Color online) (a). The transmission function in graphene nanoribbon with different defects. (b). The phonon thermal conductance v.s temperature.}
  \label{fig_trans}
\end{figure}

\begin{figure}[htpb]
  \begin{center}
    \scalebox{1.0}[1.0]{\includegraphics[width=8cm]{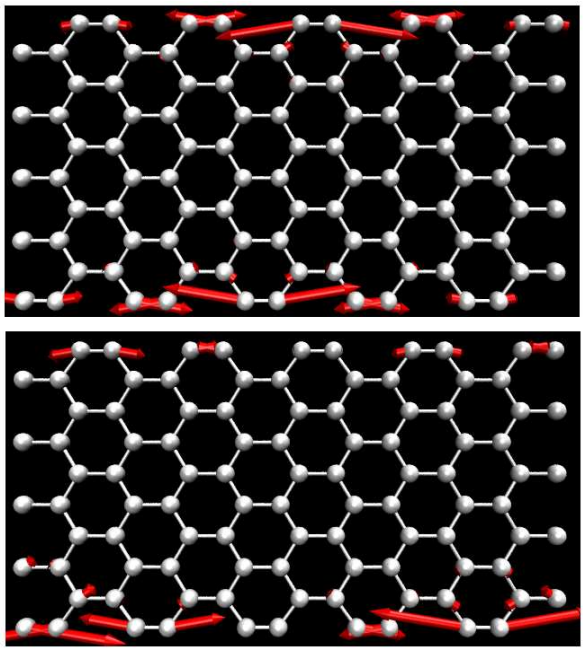}}
  \end{center}
  \caption{(Color online) The vibrational morphology of two edge modes in pure graphene nanoribbon.}
  \label{fig_umode}
\end{figure}
We first study the structure relaxation of the defect-free graphene nanoribbon as shown in Fig.~\ref{fig_cfg}~(a). We apply periodic boundary condition in the horizontal direction and the other two directions are free by introducing a vacancy space of 15~{\AA}. A larger vacancy space of 50~{\AA} has also been tried, and no obvious difference was observed. Reconstruction happens at two edges of the nanoribbon. Five different C-C bond lengths are obtained in the nanoribbon. Those ten C-C bonds at the two edges are about 1.23~{\AA} which is about $20\%$ shorter than the other inner bonds. So the edge bonds are much stronger. For the three bonds around inner carbon atoms, the horizontal one is about 1.41~{\AA} while the other two bonds are about 1.48~{\AA}. The transmission function for the defect-free graphene nanoribbon is shown in Fig.~\ref{fig_trans}~(a) by solid line (blue online). After doing correction to the $K$ matrix as discussed above, we can obtain correct value of $T[\omega=0]=3$ in low frequency region. There are some phonon modes with extremely high frequency (around 1900 cm$^{-1}$), which does not exist in two-dimensional graphene sheet. They are characteristic for the graphene nanoribbon with two edges compared with the two-dimensional graphene sheet. These modes are localized at the edges of the nanoribbon. Fig.~\ref{fig_umode} displays the vibrational property of two selective edge modes. Edge modes are originating from the stretching of edge bonds. The extremely high frequency of the edge mode is due to much stronger edge bonds compared with other inner bonds. There are totally ten edge modes in this particular graphene nanoribbon, corresponding to those ten edge C-C bonds. The thermal conductance from NEGF is shown in Fig.~\ref{fig_trans}~(b), where the value of thermal conductance at room temperature is about 1.5 nW/K. It yields a room temperature thermal conductivity of about 2800 W/m/K, taking 3.35~{\AA} as the thickness of graphene nanoribbon and 775 nm as the phonon mean free path\cite{Ghosh}. This value is close to the first-principle calculation reuslt\cite{KongBD} and that from a realistic Gruneisen parameter\cite{NikaDL}.

\begin{figure}[htpb]
  \begin{center}
    \scalebox{1.0}[1.0]{\includegraphics[width=8cm]{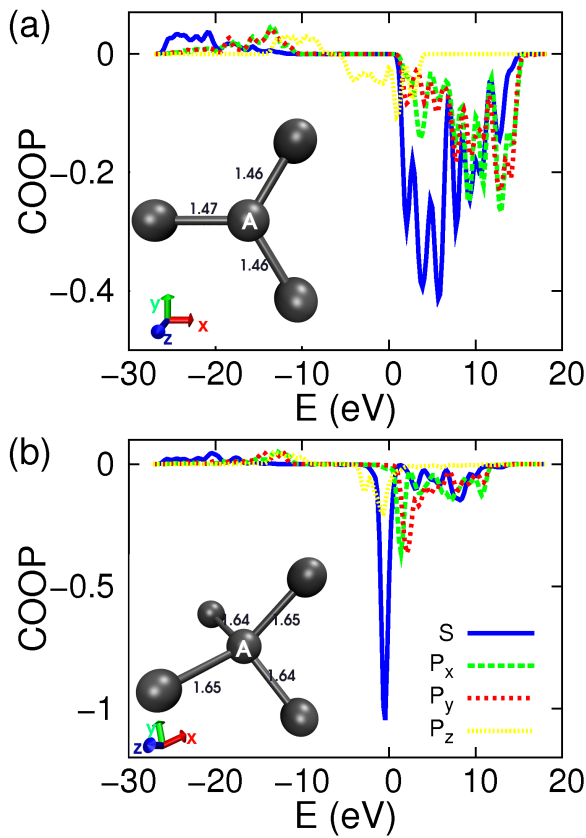}}
  \end{center}
  \caption{(Color online) The COOP in nanoribbons. (a). COOP of four electron orbitals between atom A and its three neighboring atoms in pure nanoribbon. (b). COOP of four electron orbitals between atom A in the center of vacancy defect and four neighboring atoms. Insets show the relaxed configurations.}
  \label{fig_coop}
\end{figure}

We now study the graphene nanoribbon with one vacancy defect situated at two nonequivalent positions (edge or center) as indicated by arrows in Fig.~\ref{fig_cfg}~(a). These two highlighted carbon atoms can be removed to generate a vacancy defect in the nanoribbon. Defective regions in all structures in Fig.~\ref{fig_cfg} have been highlighted by a box (green online). Fig.~\ref{fig_cfg}~(b) shows the optimized structure for nanoribbon with an edge vacancy defect. Reconstruction at the edge is realized by forming a pentagon, and only a limited region is affected. As a result, the edge vacancy defect can only introduce limited effect on the transmission function and thermal conductance shown in Fig.~\ref{fig_trans}. The reconstruction of the edge of the ribbon has direct effect on the phonon edge modes. It causes a down-shift of about 200 cm$^{-1}$ for the edge mode localized at this defective region. Fig.~\ref{fig_cfg}~(c) and (d) exhibit the relaxed configuration for the nanoribbon with a vacancy defect inner the nanoribbon. Two pentagon-hexagon 5-6 pairs have been formed around the defect. The inner vacancy defect has no effect on the phonon edge modes, as the reconstruction only takes place inner the nanoribbon. Interestingly enough, a saddle-like surface is generated around the defective region with the defective center as the saddle point. A curvature surface was quite common in defective graphene\cite{Lusk2008}. The crystal-orbital overlap population (COOP)\cite{HughbanksT} between the defective central carbon atom and its four neighboring atoms is shown in Fig.~\ref{fig_coop}~(b). The COOP for the bonding around the vacancy defect is sharper and narrower than the SP2 bonding in a defect-free nanoribbon as shown in panel (a) in the same figure. The bond length of the four defective bonds is about 1.64~{\AA}, which is even larger than the C-C bond length of 1.54~{\AA} in SP3 bonding diamond. The saddle-like surface can efficiently scatter all phonon modes in whole frequency region as shown in Fig.~\ref{fig_trans}~(a) by dash dot line (black online). The high-frequency phonon modes are scattered more seriously by this saddle surface compared with other types of defects. Only in this inner vacancy defect, the low frequency phonon mode will be considerably scattered. As a result of the strong scattering of all phonon modes by this saddle-like surface, the phonon thermal conductance is greatly reduced by the inner vacancy defect as shown in Fig.~\ref{fig_trans}~(b) by dot dash line (black online). The thermal conductance can be reduced by as much as $60\%$ at high temperatures. Considerable reduction is also observed in low-temperature region, since low-frequency phonon modes are also scattered by the inner vacancy defect.

After investigation of the vacancy defect, we now switch to the discussion of substitutional silicon defect in nanoribbon. We consider the two different positions (edge/center) as indicated by the arrows in Fig.~\ref{fig_cfg}~(a). Fig.~\ref{fig_cfg}~(e) shows the relaxed structure for Si defect at the edge of the nanoribbon. We find that the optimization is accomplished mainly in two steps. Firstly, the single Si atom will be rejected outside the graphene nanoribbon and carbon atoms around the defect undergoes reconstruction to form a pentagon similar as edge vacancy defect in panel (b). Secondly, the lonely Si atom seeks a stable position through forming a pentagon with other four carbon atoms. In this sense, the edge Si defect can be understood as the adsorption of one Si atom to the relax configuration of an edge vacancy defect. So the edge Si defect can provide a slightly stronger scattering to phonon modes than the edge vacancy defect, which is confirmed in the transmission function shown in Fig.~\ref{fig_trans}~(a) by dot line (red online). As a result, the thermal conductance is reduced a bit more by edge Si defect as shown in Fig.~\ref{fig_trans}~(b). Similarly as the edge vacancy defect, one edge C-C bond was destroyed by edge Si defect. As a result, the edge mode localized around this bond will show red-shift by the reconstruction of the edge. Fig.~\ref{fig_cfg}~(f) shows the relaxed structure for a Si defect inner the graphene nanoribbon. No curvature is generated by this point defect. The three bonds around the Si atom are elongated to be 1.7~{\AA}. The size of the Si defective region is about 3.5~{\AA}. Fig.~\ref{fig_trans}~(b) reveals an interesting fact that the thermal conductance is not sensitive to the position of the Si defect. Almost a same value is obtained for thermal conductance of nanoribbon with edge and inner Si defect. It is because the silicon and carbon elements are in the same group with four valence electrons, thus no curvature is generated around Si defect. In the meantime, the Si defect only affects a limited range (within 3.5~{\AA}). Similar as the inner vacancy defect, the inner Si defect almost has no effect on the edge modes.

\section{conclusions}
To conclude, we study the phonon thermal conductance in graphene nanoribbon by combination of {\it ab initio} SIESTA package and NEGF approach. We suggest a correction to the force constant matrix to solve the conflict between long-range character of first-principle calculation and FNN property of NEGF. We first find the relaxed structure of the nanoribbon and vibrational phonon modes. Special attention is paid to the localized edge modes. We then analyze vacancy and Si point defects in the nanoribbon and find that that the edge vacancy defect has limited effect on the thermal conductance as it only causes a further reconstruction of the edge. The inner vacancy defect gives the strongest scattering for all phonon modes by forming a saddle-like surface around the defect, thus greatly reduces the thermal conductance. For the Si defect, the thermal conductance is almost the same for edge and inner defect. We find that the edge defects can introduce a strong down-shift to the phonon edge mode localized around this defective bond, since the stronger edge bond was destroyed by edge defects.

\textbf{Acknowledgements} The work is supported in part by a Faculty Research Grant of R-144-000-257-112 of National University of Singapore.

\end{document}